\documentclass[pra,twocolumn,superscriptaddress,10pt,noshowpacs]{revtex4}
\usepackage[english]{babel}
\usepackage[T1]{fontenc}
\usepackage[utf8]{inputenc}
\usepackage{graphicx,epstopdf}
\usepackage{amssymb}
\usepackage{amsmath}
\usepackage{natbib}
\usepackage[normalem]{ulem}

\usepackage{amsfonts}
\usepackage{bbm}
\usepackage{color}
\usepackage{latexsym}
\usepackage{caption}
\usepackage{subcaption}
\usepackage{times,txfonts}
\usepackage{xcolor}
\usepackage{array}

\newcommand{\beq}{\begin{equation}}
\newcommand{\eeq}{\end{equation}}
\newcommand{\bea}{\begin{eqnarray}}
\newcommand{\eea}{\end{eqnarray}}

\begin{document}

\title{Neutrino mass generation via the inverse seesaw mechanism in a $U(1)_{B-L}$ gauge extension}

\author{H. Nogueira}
\affiliation{Departamento de Física, Universidade Federal do Ceará, 60455-760, Fortaleza, Cear\'{a}, Brazil}
\email{higo@fisica.ufc.br}

\author{J. Furtado}
\affiliation{Centro de Ci\^{e}ncias e Tecnologia - CCT, Universidade Federal do Cariri, 63048-080, Juazeiro do Norte, Cear\'{a}, Brazil}
\email{job.furtado@ufca.edu.br}

\author{R. N. Costa Filho}
\affiliation{Departamento de física, Universidade Federal do Ceará, 60455-760, Fortaleza, Cear\'{a}, Brazil}
\email{rai@fisica.ufc.br}

\author{J. Leite}
\affiliation{AHEP Group, Institut de F\'{i}sica Corpuscular --
  CSIC/Universitat de Val\`{e}ncia, Parc Cient\'ific de Paterna.\\
 C/ Catedr\'atico Jos\'e Beltr\'an, 2 E-46980 Paterna (Valencia) - SPAIN}
\email{juliorafa@gmail.com}

\begin{abstract}
  We present anomaly-free solutions suitable for an inverse seesaw realization within a $U(1)_{B-L}$ extension. Implementing such a mechanism, in a phenomenologically viable way, requires the inclusion of at least four exotic fermions, assumed to be Standard Model singlets. In order to build anomaly-free models, the $B-L$ charges of these exotic fermions must satisfy two constraint equations, known as the anomaly equations. Here, we focus on solutions involving four to eight new right-handed fermions, favoring cases where all the $B-L$ charges of these fermions are rational numbers. We showed that, when considering only the right-handed fermions necessary to realize the mechanism, the solutions must have irrational charge values. By adding one more exotic singlet fermion, which does not directly enter the mechanism mass matrix, it becomes possible to find solutions with rational charges, while the addition of a second one enables a reduction of the scalar sector. On top of the two anomaly equations, a set of inequalities and additional constraint equations were added to correctly account for neutrino masses and to minimize the scalar content. So here we present sets of charges that obey the anomaly equations as well as these additional constraints. Finally, we explore the phenomenological implications of such solutions by analyzing their capacity to provide a framework for dark matter candidates, choosing one particular solution as an example.

\end{abstract}

\maketitle

\section{Introduction}

The phenomenon of neutrino oscillations indicates that neutrinos have small masses \citep{Fukuda1998,Ahmad2002,Eguchi2003}  and suggests a modification of the Standard Model of Particle Physics (SM) \citep{Mohapatra2006}. Neutrino oscillation data confirms the existence of three neutrinos and that at least two of them are massive and non-degenerate. The correct neutrino mass spectrum can be generated through seesaw mechanisms \citep{ma1998}, where the small neutrino masses are suppressed by a high-energy scale $\Lambda$, usually associated with the presence of heavy exotic fields. 

We might classify seesaw mechanisms as low- or high-scale, depending on whether they leave imprints at the TeV scale and, consequently, whether they are accessible to future collider experiments \citep{boucenna2014}. The traditional high-scale seesaw mechanisms can be embedded in the framework of grand unified theories, as the high energy scales associated with them align with the energy scales of grand unification theories. In the standard type-I seesaw mechanism, for instance, the exotic fermions are expected to have masses of the order of $10^{14}$ GeV, rendering the associated phenomenology inaccessible to experiments \citep{mohapatra1980,valle1982}. Alternatively, low-scale seesaws, such as the inverse seesaw (ISS), may naturally lead to testable signatures.

In general, the inverse seesaw requires a more enlarged field content than simpler variants, such as the type-I, II and III seesaw mechanisms. The need for more fields is not necessarily a drawback since they can play important roles in solving other open questions and/or contribute for the theoretical consistency of a given model. For instance, new fermions are paramount when extending the SM gauge group as they can ensure that the anomaly coefficients are properly canceled. An established choice for an SM gauge extension is the gauge group $U(1)_{B-L}$, where $B$ and $L$ represent the baryon and lepton numbers, respectively. In this context, the anomaly equations take a specific form, with some standard solutions being well known. Depending on the properties of the charges in these solutions, the associated extra fermions can help to address different open problems, such as neutrino masses or dark matter \citep{bernal2019,gu2020,patra2016}.

In this paper, we present anomaly-free solutions suitable for an inverse seesaw realization within a $U(1)_{B-L}$ extension. Implementing such a mechanism, in a phenomenologically viable way, requires the inclusion of at least four exotic fermions, which we assume to be Standard Model singlets. In order to build anomaly-free models, the $B-L$ charges of these exotic fermions must satisfy two constraint equations, known as the anomaly equations. Here, we focus on solutions involving four to eight new right-handed fermions, favoring cases where all their $B-L$ charges are rational numbers. We showed that, when considering only the right-handed fermions necessary to realize the mechanism, the solutions must have irrational charge values. By adding one more exotic singlet fermion, which does not directly enter the mechanism mass matrix, it becomes possible to find solutions with rational charges, while the addition of a second one enables a reduction of the scalar sector. On top of the two anomaly equations, a set of inequalities and additional constraint equations were added to correctly account for neutrino masses and to minimize the scalar content. So, here we present sets of charges that obey the anomaly equations as well as these additional constraints.

This paper is organized as follows: in the next section, we discuss briefly the anomaly equations for the present context. In section \ref{sec3}, we review the inverse seesaw mechanism, while, in section \ref{sec4}, we present the inverse seesaw mechanism in a $U(1)_{B-L}$ extension. In section \ref{sec5}, we discuss in details the solutions obtained and, in section \ref{DM}, we explore the potential of one selected solution to serve as a framework for dark matter candidates, highlighting the broader phenomenological implications of these models. Finally, in section \ref{sec6}, we outline our conclusions.

\section{Anomaly equations}\label{sec2}

To extend the SM with an abelian gauge group of charges $X$, $U(1)_{X}$, one has to guarantee that all fermions, the SM and the exotic ones, transform in specific ways under the extended group $SU(3)_{C}\times SU(2)_{L} \times U(1)_{Y} \times U(1)_{X}$ in order to avoid gauge anomalies. This will result ultimately in constraint equations called anomaly equations. The solutions to these equations are sets of the $X$ charges. Each $U(1)_{X}$ gauge group added to the SM brings in a minimum number of exotic fields \citep{davi2020}. Here we will use a special case where the charge $X$ of a certain field equals its baryon number minus its lepton number, i.e., $X=B-L$. 

The extended group $SU(3)_{C}\times SU(2)_{L} \times U(1)_{Y} \times U(1)_{B-L}$ is anomalous if only the SM content is taken into account. On the other hand, the anomalies can be canceled with the addition of three right-handed neutrinos, with $B-L$ charges equal to $-1$, which play the role of the mediators of the type I seesaw mechanism, see for instance \citep{thooft1976}. Another possible solution involving the introduction of three right-handed fermions is $B-L= \{4,4,-5\}$ \citep{montero2009}.

Here, we will work with $N \leq 8$ additional right-handed fermions $\psi_{R}$ which are singlets under the SM group. In this case, the anomaly equations will constrain the values of the $B-L$ charges that each fermion $\psi_{R}$ can take since we assume that the SM field charges are fixed. Defining $B-L \equiv BL$, the two equations are

\begin{equation}
    \sum_{i=1}^{N} BL(\psi_{iR}) = -3
\end{equation}
and

\begin{equation}
    \sum_{i=1}^{N} BL(\psi_{iR})^{3} = -3.
\end{equation}

Thus we have a linear sum of the $BL$ charges of the exotic fermions as well as a sum of its cubics. The first equation comes from the $grav^{2} \: U(1)_{B-L}$ anomaly while the second equation derives from the $U(1)^{3}_{B-L}$ anomaly \citep{schwartzQFT}. All the subsequent sets of charges in this text are solutions to these equations, in order to construct anomaly-free models from those sets of exotic fermions. We will write the solutions as sets of Latin letters, e.g. $\{a,a,b,b,b,c\}$. Additionally, we will make use of the short hand notation $\{a,a,b,b,b,c\} \equiv 2a \: 3b \: 1c$, which in this example simply means a set of 6 right-handed fermions, 2 of which have the same charge $B-L$ charge $a$, 3 having the $B-L$ charge $b$ and 1 with $B-L$ charge of $c$. As previously mentioned, the $BL$ charge set $\{-1,-1,-1\}$ favors a standard type-I seesaw realization. Recently, studies have been conducted on the realization of the type-I \citep{ma2015} and type-II \citep{nanda2020} Dirac seesaws in a $U(1)_{B-L}$ extension. In this paper, we will look for solutions compatible with the realization of another mass generating mechanism: the inverse seesaw.

\section{Inverse seesaw mechanism}\label{sec3}

The $ISS$ mechanism \citep{mohapatra1986,gonzalez1989,deppisch2005} is commonly associated with the introduction of six singlet exotic fermions, which we will label the first three copies as $N_{R}$ and the other three copies as $S_{R}$. It is, however, possible to implement viable inverse seesaw mechanisms with more minimal sets of extra fermions, as shown in \citep{abada2014}. Looking to minimize these quantities, they showed that the mechanism  can also be realized in two other ways: $\# N_{R} = \# S_{R} = 2$ and $\# N_{R} = 2$, $\# S_{R} = 3$. Given a basis $\{\overline{\nu_{L}}, N_{R}, S_{R} \}$, the mechanism mass matrix is given by

\begin{equation}\label{matrizSSI}
    M^{ISS} = \begin{pmatrix}
       0 & M_{D} & 0 \\
        M_{D}^{T} & 0 & M_{NS} \\
        0 & M_{NS}^{T} & M_{SS}
    \end{pmatrix}.
\end{equation}

The general mass matrix $M^{ISS}$ and its sub-matrices $M_{D}$, $M_{NS}$ and $M_{SS}$ will have different dimensions depending on the numbers $\# N_{R}$ and $\# S_{R}$. Also, the energy scale range allowed in each sub-matrix varies slightly depending on these same numbers \citep{abada2014}. Ultimately, we have to ensure that the neutrino masses, as well as the mixing parameters, are correctly predicted, and the deviation from unitarity is within the experimentally acceptable range. Schematically, Up to first order, the heaviest among the light neutrino will have the following mass 
\begin{equation}\label{vmass}
    \nu^{ISS}_{mass} \sim m_{SS}\frac{m_{D}^{2}}{m_{NS}^{2}} \sim 0.1 \: eV
\end{equation}
and the deviation from unitarity may be measured by
\begin{equation}\label{deviation}
    \eta^{ISS} \sim \frac{m_{D}^{2}}{m_{NS}^{2}} \lesssim 0.01 .
\end{equation}

These values can be achieved, for instance, with $m_{D} \sim 10^{2}$ GeV, $m_{NS} \sim 10^{3}$ GeV ($10^{4}$ GeV) and $m_{SS} \sim 10$ eV ($1$ keV) being the energy scales associated to the matrices $M_{D}$, $M_{NS}$ and $M_{SS}$, respectively. By looking at the expression (\ref{vmass}) we see a fundamental difference in comparison to the neutrino mass in a standard type-I seesaw mechanism, $\nu^{SS}_{mass} \sim m_{D}^{2}/m^{SS}_{messengers}$ (with $m^{SS}_{messengers}$ being the energy scale associated to the masses of the exotic messengers). In the ISS, there are three energy scales involved, instead of the only two energy scales associated to the standard type-I SS. This means that, assuming natural values for the Yukawa couplings, the hierarchy between the energy scales involved in the ISS need not be as strong as in the type-I SS. An interesting consequence is that, in the ISS, the deviation from unitarity (\ref{deviation}) on the PMNS mixing matrix can get to a percent level. This, in turn, may lead to lepton flavor violation (LFV) events appearing on energy scales accessible to the LHC \citep{boucenna2014}. Lepton flavor violation processes, such as \(\mu \to e \gamma\) and \(\mu \to 3e\), are constrained by experiments like MEG~\cite{Baldini2018} and Mu3e~\cite{Blondel2014} but remain within reach of their projected sensitivities. These signatures make the ISS a prime target for probing new physics beyond the Standard Model. Beyond collider signatures, the ISS provides a natural framework for explaining neutrino oscillation data, as its low-scale structure allows for sizable mixing between active and sterile neutrinos. Recent global fits to oscillation data indicate that ISS models can accommodate the observed neutrino mass differences and mixing angles while predicting additional sterile states that could be probed by long-baseline experiments like DUNE~\cite{DUNE2020} and Hyper-Kamiokande~\cite{HyperK2018}.

The fact that the ISS mechanism can generate phenomena at relatively low energy scales leads to its classification as a low-energy SS mechanism, in contrast to the type-I SS which depends on a much higher energy scale, $m^{SS}_{messengers}$, and thus is classified as a high-scale SS mechanism. Furthermore, the ISS mechanism might be associated with the existence of extra light fermion gauge singlet states whose existence are suggested by recent anomalies \citep{mueller2011,arevalo2013,giunti2011}. These light states emerge from the diagonalization of the full neutral fermion mass matrix in cases where $\# N_{R} < \# S_{R}$, a scenario that will be addressed in the set of solutions discussed in this work. On the other hand, when $\#N_{R} = \#S_{R}$, the mechanism produces heavy pseudo-Dirac states with mass $\sim m_{NS}$ and a mass splitting of order $\sim m_{SS}$. In Section~\ref{DM}, we will discuss how these states can combine with the other exotic fields $C_{R}$ and $D_{R}$ to form dark matter components.

Although phenomenologically interesting, the ISS mechanism has a clear disadvantage in comparison to the standard SS. The minimum field content required to realize the ISS is twice the minimum number of fermions necessary to realize the standard SS (which are four and two, respectively \citep{abada2014}). Therefore, the ISS realization is best suited in a context where the SM is extended to include a considerable number of new fermion fields. In the next section we will study the conditions to realize the ISS mechanism within the framework of a $U(1)_{B-L}$ extension.

\section{Inverse seesaw mechanism in a $U(1)_{B-L}$ extension}\label{sec4}

In section 2, we saw that the addition of exotic fermions through an anomaly-free $U(1)_{B-L}$ extension requires the $BL$ charges of these fermions to satisfy certain constraint equations, known as the anomaly equations. In section 3, we showed that the inverse seesaw realization is also connected to the addition of exotic fermions. A natural question arises: how can we choose solutions that not only allow the ISS mechanism to occur but also minimize the fermion and/or scalar sectors?

A first approach could be to require that the $BL$ charges of each $N_{R}$ and each $S_{R}$ are equal among them. Defining $a$ as the $BL$ charge of the fermion $N_{R}$, such that $BL(N_{R}) = a$, and $b$ as the $BL$ charge of the fermion $S_{R}$, such that $BL(S_{R}) = b$, this means we should look for solutions like $\{a,a,b,b\}$, $\{a,a,b,b,b\}$, and $\{a,a,a,b,b,b\}$. This requirement is an early attempt to minimize the scalar sector, because otherwise, we would have solutions with different charges: $\{a_1,a_2,b_1,b_2\}$, $\{a_1,a_2,b_1,b_2,b_3\}$, and $\{a_1,a_2,a_3,b_1,b_2,b_3\}$. These kinds of solutions would require a larger set of scalar fields to ensure that the operators leading to the mass terms in matrix [\ref{matrizSSI}] are invariant.

The anomaly equations will take a simplified form under the assumption that all fermions of a given type have equal charges. To begin with, the anomaly equations related to a $2a : 2b$ solution will look like

\begin{equation}\label{firstabequations}
    2a + 2b = -3, \hspace{0,5cm} 2a^{3} + 2b^{3} = -3
\end{equation}
while the anomaly equations of the $2a \: 3b$ and $3a \: 3b$ solutions will assume, respectively, the following forms

\begin{equation}
    2a + 3b = -3, \hspace{0,5cm} 2a^{3} + 3b^{3} = -3
\end{equation}
and

\begin{equation}\label{lastabequations}
    3a + 3b = -3, \hspace{0,5cm} 3a^{3} + 3b^{3} = -3.
\end{equation}

These are systems of equations where the number of equations matches the number of variables. As we shall see in the next section, there is no set of rational charges which is solution to these cases. Rational solutions only arise with the addition of an extra singlet fermion, say $C_{R}$. Now, defining $BL(C_{R}) = c$, the three new systems of equations are:

\begin{equation}\label{firstabcequations}
    2a + 2b + c = -3, \hspace{0,5cm} 2a^{3} + 2b^{3} + c^{3} = -3,
\end{equation}

\begin{equation}
    2a + 3b + c = -3, \hspace{0,5cm} 2a^{3} + 3b^{3} + c^{3} = -3,
\end{equation}
and

\begin{equation}\label{lastabcequations}
    3a + 3b + c = -3, \hspace{0,5cm} 3a^{3} + 3b^{3} + c^{3} = -3.
\end{equation}

Going further, a fourth singlet fermion, $D_{R}$, could be added, contributing a charge $BL(D_{R}) = d$ to the linear equations and $d^{3}$ to the cubic equations. This addition will result in three more pairs of anomaly equations. In what follows, we will work with those nine pairs of anomaly equations which are related to the nine solution-types below:

\begin{equation}
   \begin{split}
        2a \: 2b, \hspace{0,3cm} 2a \: 3b, \hspace{0,3cm} 3a \: 3b, \hspace{1,0cm}\\
        2a \: 2b \: 1c, \hspace{0,3cm}
        2a \: 3b \: 1c, \hspace{0,3cm} 3a \: 3b \: 1c, \hspace{0,5cm}\\ 2a \: 2b \: 1c \: 1d, \hspace{0,3cm}
        2a \: 3b \: 1c \: 1d, \hspace{0,3cm} 3a \: 3b \: 1c \: 1d.
   \end{split}
\end{equation}

In each case the relevant charges are assumed to be different from zero. Also, depending on the equations and inequalities to be obeyed, there may be solutions which differ from each other by the exchanges $a\leftrightarrow b$ and/or $c \leftrightarrow d$. If one solution differs from another by these exchanges, we will choose to portrait only one of these solutions.  

The addition of the extra fields $C_{R}$ and $D_{R}$ allows for solutions with diverse properties. In these domains there are solutions with rational charges, as well as solutions where no doublet scalars are necessary. However, the expansion of the fermion sector comes also with potential problems. In this context, a given solution may accidentally allow a mass term mixing the mechanism fields $\{\nu_{L}^{c},\: N_{R}, \: S_{R} \}$ and the extra fields $\{C_{R}, D_{R}\}$. Consequently, the matrix [\ref{matrizSSI}] will be deformed and this deformation could affect the correct prediction of neutrino masses and oscillation parameters.
Those undesired mass terms might be avoided by adding extra symmetries or by using the new abelian symmetry we are working with, the $U(1)_{B-L}$ symmetry. In the next section, we will present solutions that not only satisfy the nine pairs of anomaly equations but are favorable for constructing minimal models.

\section{Solutions}\label{sec5}

In this section, we will present solutions to the pairs of anomaly equations shown previously. We classify them by the number of different variables they contain. There are three pairs with two variables, three pairs with three variables and three pairs with four variables. In the last six cases, the extra degrees of freedom can pose a problem, as they might accidentally allow for the existence of mass terms that could ultimately deform the ISS mass matrix. Nonetheless, we will show solutions in which no unnecessary mass terms are created, these solutions satisfy a set of inequalities that prevent the extra mass terms. Furthermore, with the inclusion of new fermions $C_{R}$ and $D_{R}$, it is natural to assume that new scalar fields are required. Generally, this is true, but we have found solutions where the scalar sector remains equivalent to the two-variable case, i.e., where the only exotic fermions are $N_{R}$ and $S_{R}$. Solutions that minimize the scalar sector must satisfy additional equations beyond the anomaly equations. On top of that, we will constrain the absolute values of the $B-L$ charges, since high $|B-L|$ values typically require larger vacuum expectation values for the scalar fields responsible for breaking the \( U(1)_{B-L} \) symmetry, thereby pushing up the masses of the gauge boson \( Z' \) and the heavy neutrinos \cite{mohapatra1980}.

\subsection{Solutions with two variables}

Solutions with two variables represent the cases where the fermion sector is minimal. Here, the number of variables matches the number of anomaly equations. Considering $a \neq 0$ and $b\neq 0$, the equations (\ref{firstabequations}-\ref{lastabequations}) can be solved directly, and the results are summarized on Table \ref{twovariables}. 

\begin{table}[h!]
    \centering
    \begin{tabular}{||c | c | c | c| c||}
        \hline
        Solution & Sol. Type & a & b & $\#$Scalars \\
        \hline\hline
       Sol. 1 & 2a 2b & -$\frac{3}{4} - \frac{\sqrt{21}}{12}$ & $-\frac{3}{4} + \frac{\sqrt{21}}{12}$ & 3 \\
        \hline
        Sol. 2 & 2a 3b & $\frac{9}{5} + \frac{6\sqrt{6}}{5}$& $-\frac{11}{5} - \frac{4\sqrt{6}}{5}$ & 3\\
        \hline
        Sol. 3 & 2a 3b & $\frac{9}{5} - \frac{6\sqrt{6}}{5}$ & $-\frac{11}{5} + \frac{4\sqrt{6}}{5}$ & 3\\
        \hline
    \end{tabular}
    \caption{Solutions which minimize the fermionic sector. In this case the $BL$ charges of the new fermions $N_{R}$ and $S_{R}$ are necessarily irrational. The last column shows the number of exotic scalar fields to be added in order to reproduce the mass matrix (\ref{matrizSSI}). Also, we excluded solutions with $a=0$ or $b=0$, which would lead to different seesaw mechanisms.}
    \label{twovariables}
\end{table}

We first note that the charges $a$ and $b$ are irrational, which implies that to reproduce some of the entries in matrix (\ref{matrizSSI}), scalar fields with irrational charges must be added to the model. The scalar fields transform differently under the SM group: one is an $SU(2)_{L}$ doublet, while the other two are singlets. The two singlets ,$\chi^{SS}$ and $\chi^{NS}$, along with the doublet $\Phi^{LN}$ form operators that ultimately generate the mass matrix entries. These operators are: 

\begin{equation}
    \begin{split}
         y_{ij}^{SS} \chi^{SS} \overline{(S_{iR})^{c}} \: S_{jR}, \hspace{0,3cm} y_{kl}^{NS} \chi^{NS} \overline{(N_{kR})^{c}} \: S_{lR}, \hspace{0,3cm} y^{LN}_{m}\overline{L_{L}}\: \Phi^{LN} \: N_{mR}
    \end{split}
\end{equation}
with $i,j,k,l,m$ running from $1$ to $2$ or $3$, depending on the exotic fermion content. For instance, Sol. 1 requires a $SU(2)_{L}$ scalar doublet with $BL$ charge equals to $-\frac{1}{4}+\frac{\sqrt{21}}{12}$ in order to generate the $M^{SSI}_{12}$ entry, while the $M^{SSI}_{33}$ entry would require a scalar singlet of $BL$ charge equals to $\frac{3}{2}-\frac{\sqrt{21}}{6}$. On the other hand, the $M^{SSI}_{23}$ entry requires a singlet scalar with a rational $BL$ charge: $-\frac{3}{2}$. This last feature distinguishes the scalar sector of solution 1 from the scalar sectors of solutions 2 and 3 because in these solutions all exotic scalar fields have irrational $BL$ charges.

Lastly, we note that, considering the assumptions $a\neq 0$ and $b\neq0$, there are no $3a \: 3b$ solutions and, due to an $a\leftrightarrow b$ symmetry in equations (\ref{firstabequations}), it is possible to construct a trivial new solution by interchanging the charges $a$ and $b$ in Sol. 1. Additionally, the solutions from Table \ref{twovariables} minimize the fermionic sector while also keeping the scalar sector minimal. As mentioned previously, one could build a solution with four fermions whose $BL$ charges are $\{a1, a2, b1, b2\}$, but this would require six new scalar fields instead of three. The same applies to solutions like $\{a1, a2, b1, b2, b3\}$ and $\{a1, a2, a3, b1, b2, b3\}$, which would require even more exotic scalar fields.  

\subsection{Solutions with three variables}\label{sec3variables}

We might consider adding a new right-handed singlet fermion, $C_{R}$, with $BL(C_{R}) = c$. In this scenario, the pairs of anomaly equations (\ref{firstabcequations}-\ref{lastabcequations}) involve three variables, which means, in principle, that we now have some freedom to choose from the available solutions. On the other hand, the addition of a new field $C_{R}$ may accidentally lead to the formation of undesired operators, like $y_{m}^{NC} \chi^{NC} \overline{(N_{mR})^{c}} \: C_{R}$ or $y_{n}^{SC} \chi^{SC} \overline{(S_{nR})^{c}} \: C_{R}$. Such operators lead to terms which will deform the original ISS mass matrix. A natural question is: can we guarantee that a certain solution does not deform the matrix $M^{ISS}$? This can be implemented through the new group $U(1)_{B-L}$ and, as shown next, will result in new constraints on the $BL$ charges of the scalar and fermion fields. The number of undesired operators grows with the addition of new fields. In the following section, where we will add yet another fermion $D_{R}$, the complete formalism will be shown. For now, it should be understood that, to avoid such problematic operators, the charges must satisfy the sets of inequalities (\ref{firstinequality},\ref{secondinequality}) and (\ref{fifthinequality},\ref{sixthinequality}), see appendix.

On top of that, we would like to select solutions where there are no pairs of equal or opposite charges among the variables: $|a| \neq |b| \neq |c|$. Furthermore, we would prefer solutions where all charges are rational numbers. Overall, a set of charges which is a solution to the anomaly equations in three variables should also meet the following criteria:

\begin{enumerate}
    \item The charges must obey (\ref{firstinequality3}-\ref{lastinequality3});
    \item The charges have to be different in modulus: $|a| \neq |b| \neq |c|$;
    \item The charges may only assume rational non-zero values;
    \item Their rational values are such that the numerator $n$ runs from $-20$ to $20$ and the denominator $d$ goes from $1$ to $20$. 
\end{enumerate}

These criteria, however, exclude the vast majority of possible solutions if all of them are applied simultaneously. Before going to the solutions themselves, let us examine the general solution to the case $2a \: 2b \: 1c$. Starting with the solution to the anomaly equations only, we will write two variables in terms of the independent one. The functions $a = a(c)$ and $b = b(c)$ in the general solution are

        \begin{equation}
\begin{split}
a &= \frac{(-c + 1) \sqrt{5c^2 + 22c + 21} - (c + 3)^2}{4(c + 3)}, \\
b &= \frac{(c - 1) \sqrt{5c^2 + 22c + 21} - (c + 3)^2}{4(c + 3)}.
\end{split}
\end{equation}

Looking at these expressions helps to understand why solutions with rational charges are rare. The square roots in the functions $a(c)$ and $b(c)$ make it difficult to choose a rational value for the charge $c$ such that $a(c)$ and $b(c)$ are also rational. In fact, imposing on them the four criteria enumerated previously, there is only one viable solution: 

\begin{table}[h!]
    \centering
    \begin{tabular}{||c | c | c | c| c| c||}
        \hline
        Solution & Sol. Type & a & b & c & $\#$Scalars\\
        \hline\hline
        Sol. 4 & 2a 2b 1c & -11/15 & -16/15 & 3/5 & 4\\
        \hline
    \end{tabular}
    \caption{Solutions with three variables whose charges obey the following criteria: (1) (\ref{firstinequality},\ref{secondinequality}) and (\ref{fifthinequality},\ref{sixthinequality});\hspace{0,1cm} (2)  $|a| \neq |b| \neq |c|$; (3) $a$, $b$ and $c$ being non-zero rationals numbers; (4)  $|a|, \:|b|, \:|c| \leq 20$.}
    \label{threevariables}
\end{table}

We note that disregarding any of these criteria could increase the number of possible solutions. For example, one could take the standard solution for a type I seesaw realization, $\{ -1,-1,-1 \}$, and add to it a pair of opposite charges $\{ -1,+1\}$. The new set $\{ -1,-1,-1,-1,+1\}$ configures a $2a \: 2b \: 1c$ solution with $a=b=-1$ and $c=1$. Similarly, a pair $\{-5,+5\}$ can be added to  $\{ -4,-4,5 \}$ in order to make a $2a \: 2b \: 1c$ solution with $a=-4$, $b=5$ and $c=-5$. A non-trivial alternative would be a $2a \: 2b \: 1c$ solution where $a=b=-2/5$ and $c=-7/5$. 

The addition of a new singlet fermion may lead to some problems that did not exist in the two variables case. The main problem concerns the accidental appearance of undesired mass terms. These new terms might deform the matrix $M^{ISS}$ by changing its dimensionality, texture or hierarchy. For instance, an accidental mass term $\chi_{ac}(\overline{N_{R}})^{c}C_{R}$ could lead to a total neutral fermion matrix such that, post diagonalization, returns values different than those expected for the neutrino masses. Solutions like $\{-1,-1,-1,-1,1 \}$, $\{-4,-4,5,5,-5 \}$ and $\{-2/5,-2/5,-2/5,-2/5,-7/5 \}$ lead directly to this kind of problem. Let's examine closely the third one, $\{-2/5,-2/5,-2/5,-2/5,-7/5 \}$. To reproduce the $M^{ISS}_{23}$ entry, it is necessary to add a scalar singlet $\chi_{ab}$ with $BL$ charge $BL(\chi_{ab}) = 4/5$. Now assuming that the VEV related to $\chi_{ab}$ is supposed to lie around $10^3$ GeV, as we saw in section 3. An hierarchy problem arises in this case because the gauge symmetries allow the singlet $\chi_{ab}$ to also contribute to the $M^{ISS}_{22}$ and $M^{ISS}_{33}$ entries, which are, however, supposed to be around $10$ eV. Furthermore, once $a=b$, the gauge symmetries would allow the existence of the $M^{ISS}_{13}$ entry, thus causing a texture problem.

Solution 4, on the other hand, is less problematic. In order to realize the matrix $M^{ISS}$ with this solution, one needs to add three scalar fields: a doublet with $BL$ charge $BL(\Phi_{1a}) = - (1 + a) = -4/15$, and two singlets with $BL$ charges $BL(\chi_{ab}) = -(a+b) = 27/15$ and $BL(\chi_{bb}) = -(b+b) = 32/15$. Note that this set of scalars and their gauge charges naturally reproduce the matrix $M^{ISS}$ without deforming it in any manner. In addition, a fourth scalar is necessary to give $C_{R}$ a mass term.

Apart from the solution $\{-1,-1,-1,-1,1\}$, in all scenarios we have seen so far, it is necessary to add a new scalar doublet. This is the case because the Standard Model Higgs doublet has $BL$ charge equal to zero, meaning it can only be used in solutions where $a = -1$. Thus, the addition of an extra fermion $C_{R}$ was sufficient to generate a rational solution, but it did not give us any extra choices or possibilities to minimize the scalar sector. If we add yet another singlet fermion $D_{R}$, with $BL(D_{R}) = d$, the number of solutions increases, as do the qualitative differences among them. We will examine these new possibilities in the following subsection.

\subsection{Solutions with four variables}\label{solfourvariables}

Now, the neutral fermion content is extended with another exotic singlet, $D_{R}$. This addition naturally increases the number of anomaly-free solutions. On the other hand, it also increases the number of undesired terms that a solution may accidentally allow. To make the new fermions $C_R$ and $D_R$ massive, we may need to consider the inclusion of a new singlet scalar field, $\chi_{cd}$. We know that, with our choice of basis and charge constraints, the realization of the matrix $M^{ISS}$ requires in principle three exotic scalar fields: one as an $SU(2)_{L}$ doublet and two other as singlets. Thus, in the case with four variables, the models will require a total of four exotic scalar fields, \textit{a priori}. These exotic scalar fields may contribute to mass terms different from those in the mechanism's mass matrix and different from a $m_{cd}\overline{(C_{R})^{c}} D_{R}$ mass term. These mass terms may or may not deform the matrix $M^{ISS}$ whether or not there is a mix between the fields in the basis $\{\nu_{L}^{c},N_{R}, S_{R} \}$ and the extra fermions $\{C_{R},D_{R} \}$.

Now, we return to the question of whether it is possible to make sure that a certain solution does not deform the matrix $M^{ISS}$. Additionally, we consider what can be done to minimize the scalar sector. These two requirements can be translated into new sets of equations and inequalities. Starting with the first requirement of non-deformation of the matrix $M^{ISS}$, it is necessary to ensure that the four scalar fields do not contribute to undesired mass terms. In other words, the $BL$ charges of the scalar fields and the $BL$ charges of the fermion fields must be such that only the mass terms in $M^{ISS}$ (plus mass terms that involve only the fermions $\{C_{R},D_{R} \}$) are gauge-invariant.

To have a broader view, we will write all possible renormalizable operators that lead to mass terms, considering the fermion content in four variables. To make each operator $U(1)_{B-L}$ invariant, we assume, in principle, the presence of new scalar fields with specific $BL$ charges that cancel the fermion $BL$ charges. We may classify these operators based on whether they involve a doublet scalar $\Phi$ or a singlet scalar $\chi$:

\begin{equation}
    \mathcal{L}_{mass} = \mathcal{L}_{singlets} + \mathcal{L}_{doublets}
\end{equation}
where\\

\begin{equation}\label{singletos}
     \begin{split}
      \mathcal{L}_{singlets} = \: & y_{ij}^{SS} \chi^{SS} \overline{(S_{iR})^{c}} \: S_{jR} + y_{kl}^{NS} \chi^{NS} \overline{(N_{kR})^{c}} \: S_{lR} + \\ 
      &y^{CD} \chi^{CD} \overline{(C_{R})^{c}} \: D_{R} \: + y_{m}^{NC} \chi^{NC} \overline{(N_{mR})^{c}} \: C_{R} + \\
        &y_{n}^{SC}\chi^{SC} \: \overline{(S_{nR})^{c}} \: C_{R} + y^{CC}\chi^{CC}  \overline{(C_{R})^{c}} \: C_{R} \: + \\
        &y_{o}^{ND}\chi^{ND}  \overline{(N_{oR})^{c}} \: D_{R} + y_{p}^{SD}\chi^{SD}  \overline{(S_{pR})^{c}} \: D_{R} + \\
        &y_{qr}^{NN}\chi^{NN} \overline{(N_{qR})^{c}} \: N_{rR} \: + 
        y^{DD}\chi^{DD} \overline{(D_{R})^{c}} \: D_{R} + h.c.
    \end{split} 
\end{equation}
and 

\begin{equation}\label{doubletos}
   \begin{split}
       \mathcal{L}_{doublets} = &y^{LN}_{s}\overline{L_{L}}\: \Phi^{LN} \: N_{sR} + y_{t}^{LS}\overline{L_{L}}\: \Phi^{LS} \: S_{tR} +  \\
       &y^{LC}\overline{L_{L}}\: \Phi^{LC} \: C_{R} + 
    y^{LD}\overline{L_{L}}\: \Phi^{LD} \: D_{R} + h.c.
   \end{split}
\end{equation}
with $i,j,k,l,m,n,o,p,r,s,t$ running up from 1 to 2 or 3, and $y^{XY}$ being the relevant Yukawa couplings. Now note that only the first two terms in (\ref{singletos}), the first term in (\ref{doubletos}) and their respective conjugates enter the mass matrix (\ref{matrizSSI}). What about the other 11 mass terms in (\ref{singletos}) and (\ref{doubletos})? We might classify them by whether or not they cause problems to the ISS mass matrix. The terms $y^{CD} \chi^{CD} \overline{(C_{R})^{c}} \: D_{R}$, $ y^{CC}\chi^{CC}  \overline{(C_{R})^{c}} \: C_{R}$ and $y^{DD}\chi^{DD} \overline{(D_{R})^{c}} \: D_{R}$ do not cause any deformations, while the $y_{qr}^{NN}\chi^{NN} \: \overline{(N_{qR})^{c}} \: N_{rR}$ term might exist if the VEV of the scalar singlet $\chi^{NN}$ is similar to the $\chi^{SS}$ $vev$. On the other hand, the remaining terms must be avoided.

If we want to minimize the scalar sector, the natural choice for an operator to generate a mass term for the extra fermions is $y^{CD} \chi^{CD} \overline{(C_{R})^{c}} \: D_{R}$. Therefore, the operators we want to exist are 

\begin{equation}
    \begin{split}
        &y_{ij}^{SS} \chi^{SS} \overline{(S_{iR})^{c}} \: S_{jR}, \hspace{0,1cm} y_{kl}^{NS} \chi^{NS} \overline{(N_{kR})^{c}} \: S_{lR}, \\ &y^{LN}_{s}\overline{L_{L}}\: \Phi^{LN} \: N_{sR}, \hspace{0,1cm}
        y^{CD} \chi^{CD} \overline{(C_{R})^{c}} \: D_{R}.
    \end{split}   
\end{equation}

These are the operators with scalar fields $\{\chi^{SS}, \chi^{NS}, \Phi^{LN}, \chi^{CD} \}$, while the scalar fields related to the undesired operators are $\{\chi^{NC}, \chi^{SC}, \chi^{ND}, \chi^{SD}, \Phi^{LS}, \Phi^{LC}, \Phi^{LD} \}$. A third set,  containing scalars that are neither required nor necessarily problematic, is $\{\chi^{NN}, \chi^{CC}, \chi^{DD} \}$. To prevent the generation of undesired mass terms by the scalars in the second set, it is not enough to simply state that these scalars do not exist. Even if they are absent, one of the scalars in the first set (which have to exist) may form invariant undesired mass terms, depending on their gauge properties and, in special, their $BL$ charges. 

To prevent this kind of problem, we need to ensure that the gauge symmetries prohibit the construction of such terms. In other words, we must guarantee that the $BL$ charges of the scalars in the first set are different from the $BL$ charges that the scalars in the second set would have if they existed. By following this approach, we can avoid an accidental formation of the mass terms that deform the mechanism matrix. To forbid each of these mass terms, a specific set of inequalities must be satisfied. For instance:

\begin{itemize}
    \item To avoid $y_{n}^{SC}\chi^{SC} \: \overline{(S_{nR})^{c}} \: C_{R}$ \begin{equation}\label{firstinequality}
        \begin{split}
            |BL(\chi^{SC})| \neq |BL(\chi^{SS})| &\Rightarrow b-c \neq 0, \hspace{0,3cm} 3b+c \neq 0; \\
            |BL(\chi^{SC})| \neq |BL(\chi^{NS})| &\Rightarrow a+2b+c \neq 0, \hspace{0,3cm} a-c \neq 0; \\
             |BL(\chi^{SC})| \neq |BL(\chi^{CD})| &\Rightarrow b+2c+d \neq 0, \hspace{0,3cm} b-d \neq 0 .\\
        \end{split}
    \end{equation}
\end{itemize}
or, as another example,

\begin{itemize}
    \item To avoid $y_{t}^{LS}\overline{L_{L}}\: \Phi^{LS} \: S_{tR}$ 
    \begin{equation}
        \begin{split}
            BL(\Phi^{LS}\color{black}) \neq BL(\Phi^{LN}) &\Rightarrow a-b \neq 0.
        \end{split}
    \end{equation}
\end{itemize}

There are similar sets of inequalities related to the other five operators, which the reader may encounter in Appendix \ref{apendiceA}. Some of these inequalities repeat themselves so that we have an effective total of 17 inequalities. At first, we will look for rational solutions which satisfy all of the those inequalities and one of the following three pairs of anomaly equations  

\begin{equation}\label{firstabcdequations}
    2a + 2b + c + d = -3, \hspace{0,5cm} 2a^{3} + 2b^{3} + c^{3} + d^{3} = -3,
\end{equation}

\begin{equation}
    2a + 3b + c + d = -3, \hspace{0,5cm} 2a^{3} + 3b^{3} + c^{3} + d^{3} = -3
\end{equation}
or

\begin{equation}\label{lastabcdequations}
    3a + 3b + c + d = -3, \hspace{0,5cm} 3a^{3} + 3b^{3} + c^{3} + d^{3} = -3.
\end{equation}

In the four variables case, we will consider charges of rational numerical values with numerators ranging from $-10$ to $10$ and denominators from $1$ to $10$. In summary, we now want sets of charges that meet the following criteria:

\begin{enumerate}
    \item The charges must obey simultaneously all inequalities (\ref{firstinequality}-\ref{lastinequality});
    \item The following differences in charges modulus: $|a| \neq |b|$, $|a| \neq |c|$, $|a| \neq |d|$, $|b| \neq |c|$, $|b| \neq |d|$;
    \item The charges may only assume rational non-zero values;
    \item Their rational values are such that the numerator $n$ runs from $-10$ to $10$ and the denominator $d$ goes from $1$ to $10$;
     \item They have to be solutions to one of the pairs of anomaly equations (\ref{firstabcdequations}-\ref{lastabcdequations}).
\end{enumerate}

These solutions are shown in Table \ref{table2equations}. Note that they do not satisfy any additional equations that would minimize the number of required scalar fields.

\begin{table}[h!]
    \centering
    \begin{tabular}{||c | c | c | c| c| c | c||}
        \hline
        Solution & Sol. Type & a & b & c & d &$\#$Scalars\\
        \hline\hline
        Sol. 5 & 2a 2b 1c 1d & -2/5 & -8/5 & 9/5 & -4/5 & 4 \\
        \hline
        Sol. 6 & 2a 2b 1c 1d & -10/9 & -5/9 & 4/9 & -1/9  & 4 \\ 
        \hline 
        Sol. 7 & 2a 3b 1c 1d & -1 & 1/5 & -4/5 & -4/5  & 4 \\ 
        \hline 
        Sol. 8 & 2a 3b 1c 1d & -1/6 & -5/6 & 3/2 & -5/3  & 4 \\ 
        \hline
        Sol. 9 & 2a 3b 1c 1d & 1 & -7 & 8 & 8  & 4 \\ 
        \hline
        Sol. 10 & 2a 3b 1c 1d & -1/5 & -6/5 & 2 & -9/5  & 4 \\ 
        \hline
        Sol. 11 & 2a 3b 1c 1d & 8 & -7 & 1 & 1  & 4 \\ 
        \hline
        Sol. 12 & 2a 3b 1c 1d & -4/5 & 1/5 & -1 & -1  & 4 \\ 
        \hline
         Sol. 13 & 3a 3b 1c 1d & -3 & 8 & -9 & -9  & 4 \\ 
        \hline
         Sol. 14 & 3a 3b 1c 1d & -3/8 & 1/8 & -9/8 & -9/8  & 4 \\ 
        \hline
         Sol. 15 & 3a 3b 1c 1d & -8/3 & -1/3 & 3 & 3  & 4 \\ 
        \hline
    \end{tabular}
    \caption{Solutions with four variables whose charges obey the following criteria: (1) all inequalities (\ref{firstinequality}-\ref{lastinequality}); \hspace{0,1cm} (2) one of the pairs of anomaly equations (\ref{firstabcdequations}-\ref{lastabcdequations});\hspace{0,1cm} (3) $|a| \neq |b|$, $|a| \neq |c|$, $|a| \neq |d|$, $|b| \neq |c|$, $|b| \neq |d|$;\hspace{0,1cm} (4)  $|a|, \:|b|, \:|c|, \: |d| < 15$;\hspace{0,1cm} (5) \hspace{0,1cm} $a$, $b$, $c$ and $d$ being rationals.}
    \label{table2equations}
\end{table}

Now addressing the second requirement of minimizing the scalar sector, we can: (1) substitute the scalar doublet $\Phi^{LN}$ with the conjugate Higgs doublet $\Tilde{\Phi}_{H} = i\sigma^{2}\Phi_{H}^{*}$ and (2) impose that either $\chi^{NS}$ or $\chi^{SS}$ are identical to $\chi^{CD}$. This approach would reduce the usual set of four of exotic scalars to a set of two. Each substitution adds a new constraint equation on the fermion $BL$ charges, so that instead of solving two-equation systems (\ref{firstabcdequations}-\ref{lastabcdequations}), we will now solve systems with three or four equations.

Using the conjugate Higgs doublet as a replacement for $\Phi^{LN}$ simply implies that the charge $a$ must be equal to $-1$. This substitution poses no issues once the energy scale related to the $M^{ISS}_{12}$ entry is compatible with the Higgs $vev$. The new constraint equation is then

\begin{equation}\label{higgs}
    \Phi^{LN} \rightarrow \Tilde{\Phi}_{H} \Rightarrow a+1=0.
\end{equation}

However, this equation, together with anomaly equations (\ref{firstabcdequations}-\ref{lastabcdequations}) excludes any solution where all charges are rational numbers, given the limits $|n| \leq 10 $ for the numerator and $1 \leq d \leq 10$ for the denominator. We will later revisit the substitution $\Phi^{LN} \rightarrow \Tilde{\Phi}_{H}$, but now we will study the possibilities surrounding the scalar singlets.

Regarding the exotic scalar singlets, one way to minimize the scalar sector is to replace the singlet $\chi^{CD}$ with either $\chi^{NS}$ or $\chi^{SS}$ (or their respective conjugates). Making one substitution or another will have consequences for the $m_{CD}\,\overline{(C_{R})^{c}}D_{R}$ mass term: it will be around $1$~keV or $10^4$~GeV, depending on whether we use $\chi^{SS}$ (with VEV $v_{SS} \approx 1$~keV) or $\chi^{NS}$ ($v_{NS} \approx 10^4$~GeV). Each possibility introduces two equations, as we can employ either the field or its complex conjugate.

\begin{equation}\label{substituiçãocd1}
    \begin{split}
      \chi^{CD} \rightarrow \chi^{NS} &\Rightarrow BL(\chi^{CD}) = BL(\chi^{NS})\\
        &\Rightarrow a+b-c-d=0; \\
        \chi^{CD} \rightarrow \chi^{*NS} &\Rightarrow BL(\chi^{CD}) = -BL(\chi^{NS})\\
        &\Rightarrow a+b+c+d=0;
    \end{split}
\end{equation}
and

\begin{equation}\label{substituiçãocd2}
    \begin{split}
      \chi^{CD} \rightarrow \chi^{SS} &\Rightarrow BL(\chi^{CD}) = BL(\chi^{SS})\\
        &\Rightarrow 2b-c-d=0; \\
         \chi^{CD} \rightarrow \chi^{*SS} &\Rightarrow BL(\chi^{CD}) = -BL(\chi^{SS})\\
        &\Rightarrow 2b+c+d=0.
    \end{split}
\end{equation}

New systems of equations can be formed using one of the equations above. In other words, the constraints that the charges must satisfy in this case are the five conditions listed above plus one more equation,

\begin{enumerate}
    \item The solutions should obey all criteria listed in Table \ref{table2equations}; 
    \item Additionally, they have to obey one of the equations in (\ref{substituiçãocd1}) or (\ref{substituiçãocd2}).
\end{enumerate}

We present the solutions that satisfy these criteria in Table \ref{table3equations}. In the last column we also indicate which energy scale the mass term $m_{CD}\overline{(C_{R})^{c}}D_{R}$ acquires for each solution. Note that the solutions presented below satisfy all the criteria established in Table \ref{table2equations}, as well as an additional condition given by one of the equations (\ref{substituiçãocd1}) or (\ref{substituiçãocd2}).

\begin{table}[h!]
    \centering
    \begin{tabular}{||c | c | c | c| c| c | c||}
        \hline
        Solution & Sol. Type & a & b & c & d &$m_{cd}$ energy scale\\
        \hline\hline
        Sol. 16 & 2a 2b 1c 1d & -(5/2) & -(3/2) & 3 & 2  & $\approx$ 1 keV \\
        \hline
        Sol. 17 & 2a 2b 1c 1d & 1/9 & -(10/9) & -(5/9) & -(4/9)  & $\approx$ $10^{4}$ GeV \\
        \hline
        Sol. 18 & 2a 3b 1c 1d & 4 & -(7/2) & 1 & -(4/9)  & $\approx$ $10^{4}$ GeV \\
        \hline
        Sol. 19 & 2a 3b 1c 1d & -(8/7) & -(5/7) & 1 & 3/7  & $\approx$ 1 keV \\
        \hline
        Sol. 20 & 2a 3b 1c 1d & 8/3 & -(5/3) & -(1/3) & -3  & $\approx$ 1 keV \\
        \hline
    \end{tabular}
    \caption{Solutions with four variables whose charges obey the following criteria: (1) all inequalities (\ref{firstinequality}-\ref{lastinequality}); \hspace{0,1cm} (2) one of the pairs of anomaly equations (\ref{firstabcdequations}-\ref{lastabcdequations});\hspace{0,1cm} (3) one of the equations (\ref{substituiçãocd1}-\ref{substituiçãocd2}); \hspace{0,1cm} (4) $|a| \neq |b|$, $|a| \neq |c|$, $|a| \neq |d|$, $|b| \neq |c|$, $|b| \neq |d|$;\hspace{0,1cm} (5)  $|a|, \:|b|, \:|c|, \: |d| < 10$;\hspace{0,1cm} (6) $a$ $b$, $c$ and $d$ being rationals. In the last column we indicate the energy scale associated to the mass term $m_{CD}\overline{(C_{R})^{c}}D_{R}$.}
    \label{table3equations}
\end{table}

So, in Table \ref{table2equations}, the solutions require three exotic singlet scalar fields and one exotic doublet scalar field, whereas the solutions in Table \ref{table3equations} require two exotic singlet scalar fields and one exotic doublet scalar field. We have worked with systems of two and three equations involving four variables. A natural next step is to construct systems with four equations: one pair of anomaly equations, the equation to substitute $\Phi^{LN}$ and one of the equations to substitute $\chi^{CD}$. In other words, we now want the solutions to meet all six criteria of those in Table \ref{table3equations} plus Eq. (\ref{higgs}),

\begin{enumerate}
    \item The solutions should obey all criteria listed in Table \ref{table3equations}; 
    \item Also, they have to obey Eq. (\ref{higgs}).
\end{enumerate}

Unfortunately, as we have mentioned, there are no solutions compatible with Eq. (\ref{higgs}) where all charges are rational numbers. There are, however, some special solutions where only the charges $c$ and $d$ are irrational. Moreover, the irrational parts of these charges are opposite, making their sum rational. The consequence is that all scalar fields associated with such solutions possess rational charges. These solutions are presented in Table \ref{fourvariablesminimum}.

The absence of scalar fields with irrational $B-L$ charges in the models shown in Table \ref{fourvariablesminimum} makes it impossible for $C_R$ and $D_R$ to decay into SM fields. As a result, the Dirac pair formed by $C_R$ and $D_R$, after spontaneous symmetry breaking, is stable and becomes an exciting dark matter candidate. Therefore, in such cases, the $B-L$ symmetry is not only behind the origin of tiny neutrino masses but may also provide the required stability for dark matter.

\begin{table}[h!]
    \centering
    \begin{tabular}{||c | c | c | c| c| c||}
        \hline
        Solution & Sol. Type & a & b & c & d \\
        \hline\hline
         Sol 21 & 2a 2b c d &-1 & -1/4 & $-\frac{1}{4}(1+\sqrt{10})$ & $-\frac{1}{2} + \frac{1}{4}(1+\sqrt{10})$ \\ 
 \hline
Sol 22 &2a 2b c d  &-1 & -1/4 & $-\frac{1}{4}(1-\sqrt{10})$ & $-\frac{1}{2} + \frac{1}{4}(1-\sqrt{10})$  \\ 
\hline
Sol 23 & 2a 3b c d &-1 & -1/5 & $-\frac{1}{5}(1+2\sqrt{5})$ & $-\frac{1}{5}(1-2\sqrt{5})$ \\ 
\hline
Sol 24 & 2a 3b c d  &-1 & -1/5 & $-\frac{1}{5}(1-2\sqrt{5})$ & $-\frac{1}{5}(1+2\sqrt{5})$ \\  
\hline
Sol 25 & 2a 2b c d &-1 & -2 & $\frac{1}{6}(9-\sqrt{33})$ & $3+\frac{1}{6}(-9+\sqrt{33})$ \\  
\hline
Sol 26 &2a 2b c d &-1 & -2 & $\frac{1}{6}(9+\sqrt{33})$ & $3+\frac{1}{6}(-9-\sqrt{33})$\\  
  [1ex] 
 \hline
    \end{tabular}
    \caption{Solutions with four variables whose charges obey the following criteria: (1) all inequalities (\ref{firstinequality}-\ref{lastinequality}); \hspace{0,1cm} (2) one of the pairs of anomaly equations (\ref{firstabcdequations}-\ref{lastabcdequations});\hspace{0,1cm} (3) one of the equations (\ref{substituiçãocd1}-\ref{substituiçãocd2}); \hspace{0,1cm} (4) Eq. (\ref{higgs}); \hspace{0,1cm} (5) $|a| \neq |b|$, $|a| \neq |c|$, $|a| \neq |d|$, $|b| \neq |c|$, $|b| \neq |d|$;\hspace{0,1cm} (6)  $|a|, \:|b|, \:|c|, \: |d| < 10$;\hspace{0,1cm} (7) $a$ and $b$ being rationals.}
    \label{fourvariablesminimum}
\end{table}

The exotic scalar sector of these special solutions is presented in Table \ref{scalarsectorabcd}. The second and third columns show the $BL$ charges of the singlets $\chi^{NS}$ and $\chi^{SS}$, respectively, while the last column indicates the energy scale associated to the $m_{cd}(\overline{C_{R}^{c}})D_{R}$ mass term. 

\begin{table}[h!]
\centering
\begin{tabular}{||c | c | c | c||} 
 \hline
 Solution & $\chi^{NS}$ & $\chi^{SS}$ & $m_{cd}$ energy scale \\ [0.5ex] 
 \hline\hline
 Sol. 21 & 5/4 & 1/2 & $\approx 1$ keV \\ 
 \hline
Sol. 22 & 5/4 & 1/2 & $\approx 1$ keV  \\ 
\hline
Sol. 23 & 6/5 & 2/5 & $\approx 1$ keV \\ 
\hline
Sol. 24 & 6/5 & 2/5 & $\approx 1$ keV \\  
\hline
Sol. 25 & 3 & 4 & $\approx 10^{4}$ GeV  \\  
\hline
Sol. 26 & 3 & 4 & $\approx 10^{4}$ GeV\\  
  [1ex] 
 \hline
\end{tabular}
\caption{Exotic scalar sector related to the solutions in Table \ref{fourvariablesminimum}.}
\label{scalarsectorabcd}
\end{table}

All solutions in Table \ref{fourvariablesminimum} include a total of four exotic right-handed singlet fermions, $\{N_{R}, S_{R}, C_{R}, D_{R}\}$, and two exotic scalar singlets, $\{\chi^{NS}, \chi^{SS}\}$. This represents the minimum scalar content compared to the solutions in other tables, which require three or four exotic scalar fields. On the other hand, the minimum fermion content is found in Table \ref{twovariables}, particularly in Solution 1. The process of minimizing these two sectors comes with a problem: irrational charges. In the solutions of Table \ref{fourvariablesminimum}, this issue is restricted to the charges of $C_{R}$ and $D_{R}$, which do not participate in the neutrino mass generation mechanism.

\section{Dark matter considerations}\label{DM}

The exotic particle content in our solutions varies significantly in terms of energy scale, gauge properties, and nature. While solutions such as those in Table \ref{table3equations} and Table \ref{fourvariablesminimum} fix the VEVs of their exotic scalars, other solutions, like those in Table \ref{threevariables} and Table \ref{table2equations}, allow scalars with arbitrary VEVs. Solution 4 yields a Majorana mass term for \( C_R \), whereas solutions like those in Table \ref{table3equations} and Table \ref{fourvariablesminimum} generate Dirac pairs. As a gauged extension, the gauge boson of the \( U(1)_{B-L} \) group, \( Z' \), could serve as a portal to dark matter. The states participating in the mechanism, \( N_R \) and \( S_R \), may also give rise to dark matter candidates, in the form of heavy pseudo-Dirac pairs. We now integrate the discussion of these states with the fermions \( C_R \) and \( D_R \), as well as the exotic scalar fields. We select Solution 25 as our representative case, with an exotic fermion content consisting of two copies of \( N_R \) and \( S_R \) each, along with \( C_R \) and \( D_R \). The exotic scalar content is given by two complex singlets, \( \chi^{NS} \) and \( \chi^{SS} \), whose VEVs are \( v_{NS} \sim 10^4 \, \text{GeV} \) and \( v_{SS} \sim 1 \, \text{keV} \), see Table \ref{sol25}. 

\begin{table}[h!]
\centering
\begin{tabular}{||c | c | c | c | c | c ||} 
 \hline
 Fields & \( SU(2)_L \) & \( U(1)_Y \) & \( U(1)_{B-L} \) & Copies & VEV \\ [0.5ex] 
 \hline\hline
 \( L_L \) & 2 & -1/2 & -1 & 3 & - \\ 
 \hline
 \( N_R \) & 1 & 0 & -1 & 2 & - \\ 
 \hline
 \( S_R \) & 1 & 0 & -2 & 2 & - \\ 
 \hline
 \( C_R \) & 1 & 0 & \( \frac{1}{6}(9 - \sqrt{33}) \) & 1 & - \\ 
 \hline
 \( D_R \) & 1 & 0 & \( 3 + \frac{1}{6}(-9 + \sqrt{33}) \) & 1 & - \\  
 \hline
 \( \Phi_H \) & 2 & 1/2 & 0 & 1 & \(  246 \, \text{GeV} \) \\  
 \hline
 \( \chi^{NS} \) & 1 & 0 & 3 & 1 & \(  10^4 \, \text{GeV} \) \\  
 \hline
 \( \chi^{SS} \) & 1 & 0 & 4 & 1 & \(  1 \, \text{keV} \) \\  [1ex] 
 \hline
\end{tabular}
\caption{Charge assignments of both SM and exotic fields related to Solution 25.}
\label{sol25}
\end{table}

These charge assignments prevent tree-level mixing between \( Z \) and \( Z' \), because \( \Phi_H \) carries no \( B-L \) charge and \( \chi^{NS} \) and \( \chi^{SS} \) are neutral under \( SU(2)_L \times U(1)_Y \). The scalar potential, invariant under \( SU(2)_L \times U(1)_Y \times U(1)_{B-L} \), is:

\begin{align}\label{scalar25}
V_{25} = &\ \mu_H^2 \Phi_H^\dagger \Phi_H + \lambda_H (\Phi_H^\dagger \Phi_H)^2 + \mu_{NS}^2 \chi^{NS\dagger} \chi^{NS} \notag \\
& + \lambda_{NS} (\chi^{NS\dagger} \chi^{NS})^2 + \mu_{SS}^2 \chi^{SS\dagger} \chi^{SS} + \lambda_{SS} (\chi^{SS\dagger} \chi^{SS})^2 \notag \\
& + \lambda_{H-NS} (\Phi_H^\dagger \Phi_H)(\chi^{NS\dagger} \chi^{NS}) + \lambda_{H-SS} (\Phi_H^\dagger \Phi_H)(\chi^{SS\dagger} \chi^{SS}) \notag \\
& + \lambda_{NS-SS} (\chi^{NS\dagger} \chi^{NS})(\chi^{SS\dagger} \chi^{SS}),
\end{align}
with \( \mu_H^2, \mu_{NS}^2, \mu_{SS}^2 < 0 \) and \( \lambda_H, \lambda_{NS}, \lambda_{SS} > 0 \), whereas \( \lambda_{H-NS}, \lambda_{H-SS}, \lambda_{NS-SS} \) are real and typically small. During symmetry breaking, the fields acquire VEVs:
\begin{itemize}
    \item \( \Phi_H = \begin{pmatrix} 0 \\ \frac{v_H + h}{\sqrt{2}} \end{pmatrix} \), \( v_H \approx 246 \, \text{GeV} \),
    \item \( \chi^{NS} = \frac{v_{NS} + \sigma_{NS} + i \pi_{NS}}{\sqrt{2}} \), \( v_{NS} \sim 10^4 \, \text{GeV} \),
    \item \( \chi^{SS} = \frac{v_{SS} + \sigma_{SS} + i \pi_{SS}}{\sqrt{2}} \), \( v_{SS} \sim 1 \, \text{keV} \).
\end{itemize}

By substituting these expressions into (\ref{scalar25}) and computing the second derivatives of the real fields at the vacuum (\( h = \sigma_{NS} = \sigma_{SS} = 0 \)), we obtain the \( 3 \times 3 \) scalar mass matrix for the real components. In the basis \( (h, \sigma_{NS}, \sigma_{SS}) \), this matrix is written as:

\begin{equation}\label{matrizsol25}
    \begin{pmatrix}
        2 \lambda_H v_H^2 & \lambda_{H-NS} v_H v_{NS} & \lambda_{H-SS} v_H v_{SS} \\
        \lambda_{H-NS} v_H v_{NS} & 2 \lambda_{NS} v_{NS}^2 & \lambda_{NS-SS} v_{NS} v_{SS} \\
        \lambda_{H-SS} v_H v_{SS} & \lambda_{NS-SS} v_{NS} v_{SS} & 2 \lambda_{SS} v_{SS}^2
    \end{pmatrix}.
\end{equation}

Note that the upper \( 2 \times 2 \) block concentrates the high-energy scales, and its eigenvalues receive only small corrections from off-diagonal terms. We denote the three physical real scalar states as \( h_1 \), \( h_2 \), and \( h_3 \), arising predominantly from \( h \), \( \sigma_{NS} \), and \( \sigma_{SS} \), respectively. The real scalar \( h_1 \) is identified with the SM Higgs field, and its mass term receives small corrections from (\ref{matrizsol25}):

\begin{equation}
    m_{h_1}^2 \approx 2 \lambda_H v_H^2 + \frac{(\lambda_{H-NS} v_H v_{NS})^2}{2 \lambda_{NS} v_{NS}^2 - 2 \lambda_H v_H^2} + \frac{(\lambda_{H-SS} v_H v_{SS})^2}{2 \lambda_H v_H^2 - 2 \lambda_{SS} v_{SS}^2}.
\end{equation}

In the pseudoscalar sector, $\pi_{NS}$ and $\pi_{SS}$ mix to form the new states $G_{Z'}$ and $A$. After normalization, these states are given by 

\begin{equation}
G_{Z'} = \frac{3 v_{NS} \pi_{NS} + 4 v_{SS} \pi_{SS}}{\sqrt{9 v_{NS}^2 + 16 v_{SS}^2}}, \hspace{0,5cm} A = \frac{-4 v_{SS} \pi_{NS} + 3 v_{NS} \pi_{SS}}{\sqrt{9 v_{NS}^2 + 16 v_{SS}^2}}.
\end{equation}

The $G_{Z'}$ mode is absorbed by the gauge boson $Z'$, while $A$ remains a massless physical state. This massless state can be avoided by adding to the scalar potential (Eq.~(27)) the non-renormalizable operator $V_{\text{break}} = \frac{\lambda_{\text{break}}}{\Lambda_{\text{break}}^3} (\chi^{NS})^4 (\chi^{SS\dagger})^3 + \text{h.c.}$ (solutions 23 and 24 allow for renormalizable non-Hermitian operators),
which explicitly breaks an inherent global symmetry and generates a mass term for $A$:

\begin{equation}
    m^{2}_{A} = \frac{9 |\lambda_{break}| v_{NS}^4 v_{SS}}{2^{5/2} \Lambda_{break}^{3}}
\end{equation}
by choosing $\Lambda_{break} = 5.4 \times 10^{10}$ and $\lambda_{break} = 1$, we have $m_{a} \approx 10^{-2}$ eV. 

The parameter space must satisfy constraints that maintain the Higgs mass value, \( m_{h_1} \approx 125 \, \text{GeV} \), and produce appropriate masses for other states. We adopt the following parameters:
\begin{itemize}
    \item Higgs sector: \( v_H = 246 \, \text{GeV} \), \( 2 \lambda_H v_H^2 = (125 \, \text{GeV})^2 \Rightarrow \lambda_H \approx 0.129 \),
    \item Heavy scalar: \( v_{NS} = 10^4 \, \text{GeV} \), \( \lambda_{NS} = 0.1 \),
    \item Light scalar: \( v_{SS} = 1 \, \text{keV} \), \( \lambda_{SS} = 0.5 \),
    \item Portal couplings: \( \lambda_{H-NS} = 8 \times 10^{-5} \), \( \lambda_{H-SS} = 5 \times 10^{-20} \), \( \lambda_{NS-SS} = 5\times 10^{-20} \).
\end{itemize}

These values yield the following scalar states:
\begin{itemize}
    \item \( h_1 \) with mass \( m_{h_1} \approx 125 \, \text{GeV} \): identified as the SM-like Higgs boson (approximately 99\% \( h \)-component),
    \item \( h_2 \) with mass \( m_{h_2} \approx 4472 \, \text{GeV} \): dominantly composed of \( \sigma_{NS} \),
    \item \( h_3 \) with mass \( m_{h_3} \approx 1 \, \text{keV} \): essentially pure \( \sigma_{SS} \).
\end{itemize}

Within the fermion sector, the fields \( C_R \) and \( D_R \) form a Dirac fermion \( \psi_{CD} \) with mass \( m_{\psi_{CD}} = \frac{y^{CD} v_{NS}}{\sqrt{2}} \approx 12.5 \, \text{TeV} \) (\( y^{CD} = 1.77 \)). Furthermore, the heavy sector of the seesaw mechanism produces pseudo-Dirac pairs \( \psi_{NS\pm} \) with masses \( m_{\psi_{NS\pm}} = \frac{y^{NS} v_{NS}}{\sqrt{2}} \pm \frac{y^{SS} v_{SS}}{\sqrt{2}} \approx 4455 \, \text{GeV} \pm 0.707 \, \text{keV} \) (\( y^{NS} \approx 0.63 \), \( y^{SS} \approx 1 \)).

Having established the particle spectrum of Solution 25, we now qualitatively explore the potential of its exotic fields—\( \psi_{CD} \), \( \psi_{NS\pm} \), \( h_2 \), and \( h_3 \), alongside the \( Z' \) gauge boson—to account for dark matter. This framework offers a variety of candidates that could contribute to the cosmic dark matter density through distinct mechanisms, such as thermal freeze-out, coannihilation, or non-thermal production. We assess their alignment with current experimental data and their testability in future observations, without prescribing a definitive multi-component model.

\begin{equation}\label{eq:omega_total}
\Omega_{\text{DM}} \approx \underbrace{\Omega_{\psi_{CD}}}_{\text{WIMP}} + \underbrace{\Omega_{\psi_{NS\pm}} + \Omega_{h_2}}_{\text{Coannihilation}} + \underbrace{\Omega_{h_3}}_{\text{Light scalar}} + \underbrace{\Omega_{A}}_{\text{Axion-like particle}}.
\end{equation}

The Dirac fermion \( \psi_{CD} \), with a mass of approximately 12.5 TeV, is a compelling WIMP candidate due to its interactions mediated by the \( Z' \) boson \cite{Mohapatra1980}. Its stability arises from the non-integer \( B-L \) charges of \( C_R \) and \( D_R \), which forbid tree-level decays to Standard Model (SM) particles, or from an optional \( Z_2 \) symmetry that ensures no decays occur. In the early universe, \( \psi_{CD} \) could achieve thermal freeze-out through \( Z' \)-mediated annihilation processes, such as \( \psi_{CD} \overline{\psi_{CD}} \to f \overline{f} \), producing SM fermions or bosons \cite{Aprile2018, Akerib2022, Ackermann2015, Aalbers2023}. Direct detection experiments, such as XENON1T and LZ, impose tight constraints on WIMP-nucleus scattering for masses above 1 TeV, implying that \( \psi_{CD} \)'s coupling to the SM must be mediated by a heavy portal to remain viable \cite{Aprile2018, Akerib2022}. Indirect detection searches, like those conducted by Fermi-LAT for gamma-ray signals from WIMP annihilation, further restrict heavy WIMPs but permit models with low annihilation rates \cite{Ackermann2015}. Looking ahead, the DARWIN experiment could extend sensitivity to multi-TeV WIMPs, potentially probing a \( \psi_{CD} \)-like candidate if its interactions are sufficiently strong \cite{Aalbers2023}. Also, we would like to highlight the freedom associated with the parameter $y^{CD}$, which can be used to adjust the mass of $\psi_{CD}$ and prevent overabundance (by tuning it to the $Z'$ resonance, for instance).

The pseudo-Dirac fermions \( \psi_{NS\pm} \), alongside the heavy scalar \( h_2 \), present a compelling case for coannihilation. Their near-degenerate masses, both tied to the \( \chi^{NS} \) VEV, enable efficient coannihilation in the early universe, where \( \psi_{NS\pm} \) and \( h_2 \) might collectively annihilate into SM particles or \( Z' \)-mediated states \cite{Griest1991}. Stability for $\psi_{NS\pm}$ could arise from a dedicated symmetry, or from the small mass splitting which ensures any allowed decays occur on cosmological timescales, effectively making them stable for DM purposes. Similarly, \( h_2 \) requires a stabilizing mechanism, such as a \( Z_2 \) symmetry (\( \sigma_{NS} \to -\sigma_{NS} \)), to suppress decays like \( h_2 \to h_1 h_1 \), which could be induced by the portal coupling \( \lambda_{H-NS} \). LHC searches by ATLAS and CMS constrain heavy scalars and fermions in \( B-L \) models, setting lower bounds on \( Z' \) masses around 4--5 TeV for couplings like \( g_{B-L} \approx 0.57 \) \cite{ATLAS2021, CMS2022}. The High-Luminosity LHC (HL-LHC) and future colliders, such as the Future Circular Collider (FCC-hh), could produce \( \psi_{NS\pm} \) or \( h_2 \) in association with \( Z' \), providing direct tests of their DM candidacy \cite{Golling2017}.

The light scalar \( h_3 \) emerges as a intriguing non-thermal DM candidate. It originates from \( \chi^{SS} \), benefiting from an extremely small mixing with the SM Higgs, which ensures stability over cosmological timescales. This scalar could be produced via the misalignment mechanism \cite{Preskill1983}, where early-universe field displacements lead to coherent oscillations that contribute to the DM density. Its keV-scale mass makes direct detection challenging, as its interactions with SM particles are highly suppressed. However, cosmological observations, such as Cosmic Microwave Background (CMB) measurements from Planck, constrain the relic density of light scalars, necessitating careful tuning of initial conditions to avoid overproduction \cite{Planck2018}. Future experiments, like the Simons Observatory, could detect signatures of fields like \( h_3 \) through CMB spectral distortions, offering sensitivity to keV-scale relics \cite{Chluba2021}. 

The $A$ state, with mass $\sim 10^{-2}$ eV, naturally fits the profile of an axion-like particle (ALP) and can be produced through the misalignment mechanism in the early universe. Its decay constant is given by
$f_A \sim \sqrt{9 v_{NS}^2 + 16 v_{SS}^2} \approx 3 v_{NS} \sim 3 \times 10^4 \, \text{GeV}$. ALPs with a mass of \( \sim 10^{-2} \, \text{eV} \) are subject to astrophysical constraints, as they can be produced in stellar environments. The $A$ state could provide a small contribution to the dark matter relic density, provided its coupling to photons complies with astrophysical bounds, namely \( g_{A\gamma} \sim 10^{-10} \, \text{GeV}^{-1} \). An ALP at this energy scale could be probed in future experiments such as IAXO, which is planned for the coming years \cite{Armengaud2014,IAXO2025}.

The \( Z' \) boson, with a mass of approximately 5.7 TeV and gauge coupling \( g_{B-L} \approx 0.57 \), acts as a pivotal portal linking DM candidates to the SM. Its heavy mass aligns with current LHC constraints, which require high-luminosity runs or next-generation colliders for direct production \cite{ATLAS2021}. The absence of significant \( Z-Z' \) mixing, ensured by Solution 25’s charge assignments, is consistent with electroweak precision measurements, bolstering the model’s viability \cite{Appelquist2003}. 

\begin{table}[h!]
\centering
\begin{tabular}{||c | c | c | p{2cm}||} 
 \hline
 Fields & Mass & Nature &  DM role \\ [0.5ex] 
 \hline\hline
 $h_{2}$ & 4472 GeV & $\sigma_{NS}$-like & Coannihilation partner \\ 
 \hline
$h_{3}$ & 1 keV & $\sigma_{SS}$-like & Light scalar DM  \\ 
\hline
$A$ & $10^{-2}$ eV & Axion-like & ALP \\ 
\hline
$\psi_{CD}$ & 12.5 TeV & Dirac fermion & Thermal WIMP \\  
\hline
$\psi_{NS\pm}$ & 4455 GeV $\pm$ 0.707 keV & Pseudo-Dirac & Coannihilation  \\  
\hline
$Z'$ & 5.7 TeV & Gauge boson & Mediator \\  
  [1ex] 
 \hline
\end{tabular}
\caption{Dark matter components arising from Solution 25, with masses calculated at tree level. Other solutions may lead to very different dark matter candidates, varying in mass, nature, and role.}
\label{DMcandidates}
\end{table}

The interplay of \( \psi_{CD} \)'s WIMP-like freeze-out, the coannihilation dynamics of \( \psi_{NS\pm} \) and \( h_2 \), and the non-thermal production of \( h_3 \) and the axion-like particle $A$ offers a diverse phenomenological landscape, as summarized in Table \ref{DMcandidates}. While achieving the observed relic density (\( \Omega_{DM} h^2 \approx 0.12 \)) may require a combination of these mechanisms and a detailed exploration of the parameter space, the distinct signatures of these candidates offer multiple avenues for experimental scrutiny. Furthermore, it is worth emphasizing that this solution represents just one option among the diverse set of anomaly-free solutions explored in this work.
The variety of solutions, differing in their fermion and scalar content as well as their $ B-L $ charge assignments, suggests that other configurations could yield distinct dark matter candidates, potentially ranging from lighter or heavier WIMPs to scalar states with different production mechanisms. These alternative candidates may offer unique advantages alongside potential challenges compared to those of Solution 25. This diversity invites further investigation into the phenomenological implications of such solutions.

\section{Conclusion}\label{sec6}

The inverse seesaw mechanism requires a larger set of exotic fermions compared to standard seesaw mechanisms. Therefore, its realization within a $U(1)_{B-L}$ extension of the Standard Model seems natural. In this work, we explored anomaly-free solutions for the ISS framework, prioritizing solutions where all exotic fermions possess rational $B-L$ charges. In the search for such solutions, we identified the need to include additional fermions -- exotic fermions that do not directly participate in the seesaw mechanism. This inclusion allowed for solutions where all fermions possess rational $B-L$ charges. In order to preserve the ISS mass matrix, we wrote constraints, in the form of inequalities, that the $B-L$ charges must satisfy to prevent undesired operators. Additionally, we introduced other charge constraints, in the form of equations, ensuring that the anomaly-free solutions complying with them would result in minimal scalar sectors. All solutions are equally efficient in generating the correct mass matrix and, consequently, in predicting the correct neutrino masses. However, the particle content differs across solutions, affecting both the fermion and scalar sectors, along with their gauge properties, which in turn determine the operators permitted by each solution. Among the various cases we explored, we identified some in which the extra fermions required to cancel the $B-L$ anomalies are stable and could serve as dark matter candidates. To this end, we presented in the last section a qualitative analysis of one such solution (Solution 25), exploring its potential as a unified framework for neutrino mass generation and dark matter phenomenology. We identified a set of exotic fields as dark matter candidates, assessing their production mechanisms and experimental signatures, though further study is needed to confirm their contributions to the observed relic density.  Moreover, the plurality across solutions invites individual analyses of each framework as a potential source of dark matter candidates.

\appendix
\section{Inequalities to be obeyed in order to avoid extra operators}\label{apendiceA}

\subsection{Three variables}\label{inequalities3}
The addition of an exotic fermion $C_{R}$ may lead to operators that can deform the general neutral fermion mass matrix. Here, we establish a set of inequalities that the $B-L$ charges of the relevant fields must satisfy to prevent such problematic operators in the three-variable case (Sec. \ref{sec3variables}). These are: 

\begin{itemize}
    \item To avoid $y_{n}^{SC} \chi^{SC} \: \overline{(S_{nR})^{c}} \: C_{R}$ \begin{equation}\label{firstinequality3}
        \begin{split}
            |BL(\chi^{SC})| \neq |BL(\chi^{SS})| &\Rightarrow b-c \neq 0, \hspace{0,3cm} 3b+c \neq 0; \\
            |BL(\chi^{SC})| \neq |BL(\chi^{NS})| &\Rightarrow a+2b+c \neq 0, \hspace{0,3cm} a-c \neq 0; \\
             |BL(\chi^{SC})| \neq |BL(\chi^{CC})| &\Rightarrow 3b+c \neq 0, \hspace{0,3cm} b-c\neq 0 .\\
        \end{split}
    \end{equation}
     \item To avoid $y_{m}^{NC} \chi^{NC} \overline{(N_{mR})^{c}} \: C_{R}$ \begin{equation}
        \begin{split}
            |BL(\chi^{NC})| \neq |BL(\chi^{SS})| &\Rightarrow a-2b+c \neq 0, \hspace{0,3cm} a+2b+c \neq 0; \\
            |BL(\chi^{NC})| \neq |BL(\chi^{NS})| &\Rightarrow 2a+b+c \neq 0, \hspace{0,3cm} b-c \neq 0; \\
             |BL(\chi^{NC})| \neq |BL(\chi^{CC})| &\Rightarrow a+3c \neq 0, \hspace{0,3cm} a-c \neq 0  .\\
        \end{split}
        \end{equation}
        \item To avoid $y_{t}^{LS}\overline{L_{L}}\: \Phi^{LS} \: S_{tR}$ 
    \begin{equation}
        \begin{split}
            BL(\Phi^{LS}) \neq BL(\Phi^{LN}) &\Rightarrow a-b \neq 0;
        \end{split}
    \end{equation}
        \item To avoid $y^{LC}\overline{L_{L}}\: \Phi^{LC} \: C_{R}$ 
    \begin{equation}\label{lastinequality3}
        \begin{split}
            BL(\Phi^{LC}) \neq BL(\Phi^{LN}) &\Rightarrow a-c \neq 0.
        \end{split}
    \end{equation}
\end{itemize}

\subsection{Four variables}
Here we write all the inequalities mentioned in Sec.\ref{solfourvariables}. There are a total of seven operators we want to prohibit, four of them involves scalar $SU(2)_{L}$ singlets and the other three are built with scalar $SU(2)_{L}$ doublets. This would give a total of $4 \times 6 + 3 \times 1 = 27$ inequalities. However, some of these inequalities repeat themselves, resulting in a total of 17 different inequalities. For completeness, we list below all 27 inequalities, indicating to which operator they are related

\begin{itemize}
    \item To avoid $y_{n}^{SC} \chi^{SC} \: \overline{(S_{nR})^{c}} \: C_{R}$ \begin{equation}\label{firstinequality}
        \begin{split}
            |BL(\chi^{SC})| \neq |BL(\chi^{SS})| &\Rightarrow b-c \neq 0, \hspace{0,3cm} 3b+c \neq 0; \\
            |BL(\chi^{SC})| \neq |BL(\chi^{NS})| &\Rightarrow a+2b+c \neq 0, \hspace{0,3cm} a-c \neq 0; \\
             |BL(\chi^{SC})| \neq |BL(\chi^{CD})| &\Rightarrow b+2c+d \neq 0, \hspace{0,3cm} b-d \neq 0 .\\
        \end{split}
    \end{equation}
     \item To avoid $y_{m}^{NC} \chi^{NC} \overline{(N_{mR})^{c}} \: C_{R}$ \begin{equation}\label{secondinequality}
        \begin{split}
            |BL(\chi^{NC})| \neq |BL(\chi^{SS})| &\Rightarrow a-2b+c \neq 0, \hspace{0,3cm} a+2b+c \neq 0; \\
            |BL(\chi^{NC})| \neq |BL(\chi^{NS})| &\Rightarrow 2a+b+c \neq 0, \hspace{0,3cm} b-c \neq 0; \\
             |BL(\chi^{NC})| \neq |BL(\chi^{CD})| &\Rightarrow a+2c+d \neq 0, \hspace{0,3cm} a-d \neq 0 .\\
        \end{split}
        \end{equation}
        \item To avoid $y_{p}^{SD}\chi^{SD}  \overline{(S_{pR})^{c}} \: D_{R}$ \begin{equation}
        \begin{split}
            |BL(\chi^{SD})| \neq |BL(\chi^{SS})| &\Rightarrow b-d \neq 0, \hspace{0,3cm} 3b+d \neq 0; \\
            |BL(\chi^{SD})| \neq |BL(\chi^{NS})| &\Rightarrow a+2b+d \neq 0, \hspace{0,3cm} a-d \neq 0; \\
             |BL(\chi^{SD})| \neq |BL(\chi^{CD})| &\Rightarrow b+c+2d \neq 0, \hspace{0,3cm} b-c \neq 0 .\\
        \end{split}
    \end{equation}
    \item To avoid $y_{o}^{ND}\chi^{ND}  \overline{(N_{oR})^{c}} \: D_{R}$ \begin{equation}
        \begin{split}
            |BL(\chi^{ND})| \neq |BL(\chi^{SS})| &\Rightarrow a-2b+d \neq 0, \hspace{0,3cm} a+2b+d \neq 0; \\
            |BL(\chi^{ND})| \neq |BL(\chi^{NS})| &\Rightarrow 2a+b+d \neq 0, \hspace{0,3cm} b-d \neq 0; \\
             |BL(\chi^{ND})| \neq |BL(\chi^{CD})| &\Rightarrow a+c+2d \neq 0, \hspace{0,3cm} a-c \neq 0 ;\\
        \end{split}
    \end{equation}
     \item To avoid $y_{t}^{LS}\overline{L_{L}}\: \Phi^{LS} \: S_{tR}$ 
    \begin{equation}\label{fifthinequality}
        \begin{split}
            BL(\Phi^{LS}) \neq BL(\Phi^{LN}) &\Rightarrow a-b \neq 0;
        \end{split}
    \end{equation}
    \item To avoid $y^{LC}\overline{L_{L}}\: \Phi^{LC} \: C_{R}$ 
    \begin{equation}
        \begin{split}\label{sixthinequality}
            BL(\Phi^{LC}) \neq BL(\Phi^{LN}) &\Rightarrow a-c \neq 0;
        \end{split}
    \end{equation}
    \item To avoid $y^{LD}\overline{L_{L}}\: \Phi^{LD} \: D_{R}$ 
    \begin{equation}\label{lastinequality}
        \begin{split}
            BL(\Phi^{LD}) \neq BL(\Phi^{LN}) &\Rightarrow a-d \neq 0.
        \end{split}
    \end{equation}
\end{itemize}

\end{document}